**Title: Unexpected Universality in the Viscosity of Metallic Liquids**

**Authors:** M. Blodgett[1], T. Egami[2], Z. Nussinov[1*] and K. F. Kelton[1*]

**Affiliations:**

[1]Department of Physics and Institute of Materials Science and Engineering, Washington University, St. Louis, MO 63130 U.S.A.

[2]University of Tennessee, Knoxville TN and Joint Institute for Neutron Sciences, Oak Ridge National Laboratory, Oak Ridge, TN U.S.A.

*Correspondence to: kfk@physics.wustl.edu (Primary), zohar@wuphys.wustl.edu (Secondary)

**Abstract:** The range of the magnitude of the liquid viscosity, $\eta$, as a function of temperature is one of the most impressive of any physical property, changing by approximately 17 orders of magnitude from its extrapolated value at infinite temperature ($\eta_o$) to that at the glass transition. We present experimental measurements of containerlessly processed metallic liquids that reveal that $\log(\eta/\eta_o)$ as a function of $T_{coop}/T$ is a universal curve. The temperature $T_{coop}$ corresponds to the onset of cooperative motion and is strongly correlated with the glass transition temperature. On average $\eta_o$ is found to be $nh$, where $h$ is Planck's constant and $n$ is the particle number density. A surprising universality in the viscosity of metallic liquids and its relation to the glass transition is demonstrated.

**One Sentence Summary:** Experimental measurements reveal a universal scaling relation for the temperature dependent viscosity of liquid metals.

**Main Text:** The nature of the dynamical processes in liquids and how liquids transform to glasses are major outstanding questions in condensed matter science. The shear viscosity is a particularly temperature-sensitive property for glass-forming liquids, changing by about 17 orders of magnitude upon cooling from high temperatures to the glass transition temperature, $T_g$. The way in which the viscosity, or related relaxation times, change with temperature scaled to $T_g$ is the basis for the widely-used *fragility* classification scheme introduced by Angell (*1*). For very *strong* liquids the viscosity shows an Arrhenius behavior, with a well-defined activation energy over a wide temperature range that extends from above the melting temperature down to $T_g$. The viscosities of fragile liquids are characterized by activation energies that are small at high temperature and increase rapidly upon approaching $T_g$. The strongest glass-formers are network oxides, while molecular liquids such as o-terphenyl, decalin and isoquinoline, are among the most fragile. Upon close examination, some non-Arrhenius behavior is observed near the glass transition, even in strong liquids, but this becomes more dramatic as the fragility of the liquid increases. Thermodynamic (*2*) and direct structural signatures (*3*) of fragility support a connection between structure and dynamics in liquids, which has been long assumed. The concept of fragility appears to provide a coherent scheme for classifying all liquids and linking to glass formability in some cases. However, the use of the glass transition as the scaling temperature (as in an Angell plot (*1*)) can be questioned since it is somewhat arbitrary, defined as the temperature at which the viscosity reaches the value of $10^{12}$ Pa.s. The viscosity experimental data for liquid metals presented here demonstrate the existence of a high-temperature universal scaling temperature, $T_{coop}$, which has been predicted from molecular dynamics (MD) simulations of metallic liquids (*4*) and theoretical studies of non-metallic glass-forming liquids (*5-7*). Scaling the temperature, $T$, by $T_{coop}$ (a temperature that corresponds to the onset of dynamical cooperativity in the liquid), and scaling the viscosity by $\eta_o$ (the high temperature limit on the viscosity in the liquid phase) yields a universal curve that fits the viscosities of all liquid metals studied, from above the melting temperature to the glass transition temperature.

**Measurements of the Liquid Viscosity and a Scaled Universal Curve**

The viscosities of a variety of metallic liquids were measured at high temperatures in a high-vacuum containerless environment, using electrostatic levitation (see Supplementary Information). These are shown as a function of inverse temperature in Figure 1.a. The liquids studied include strong liquids that easily form metallic glasses, such as $Zr_{57}Cu_{15.4}Ni_{12.6}Al_{10}Nb_5$ (Vit 106) and $Zr_{58.5}Cu_{15.6}Ni_{12.8}Al_{10.3}Nb_{2.8}$ (Vit 106A), and more marginal glass-forming fragile liquids, such as Ti-Zr-Cu-Pd, Cu-Zr, where faster cooling rates are required for glass formation. Also included are liquids for which glass formation has not been observed, such as Ni-Si. Figure 1.b shows an Angell plot for these data, presenting log($\eta$) as a function of $T_g/T$. Scaling the temperature by $T_g$ reduces the scatter of the data from Figure 1.a, but significant variations among the data remain.

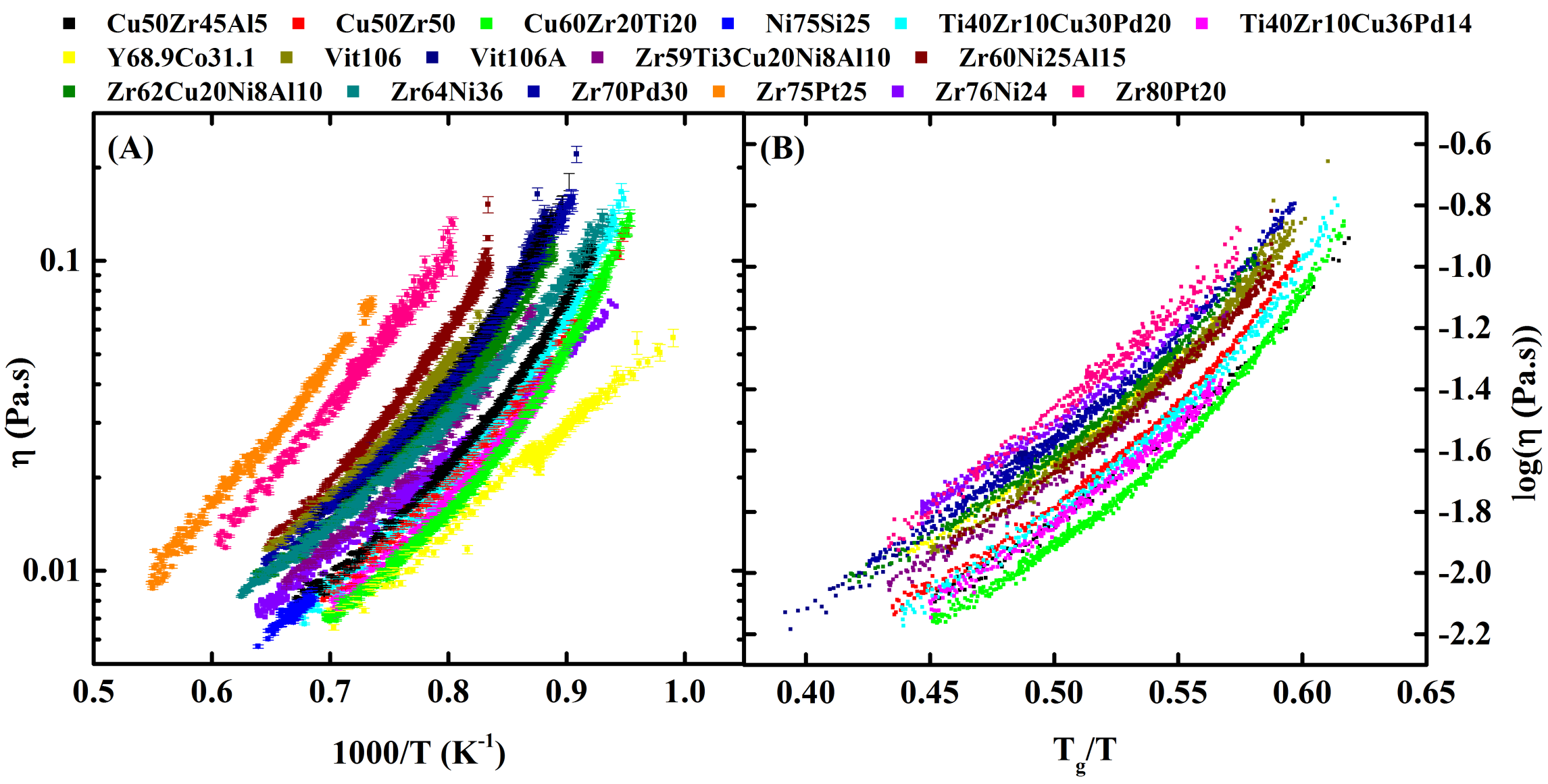

**Figure 1** – (a) Measured viscosity, η, as a function of temperature for metallic liquids. (b) An Angell plot of the log(η) as a function of $T_g/T$.

As illustrated in Figure 2.a for a $Zr_{64}Ni_{36}$ liquid, the viscosities of all of the data in Figure 1 have Arrhenius temperature dependences at a sufficiently high temperature. This agrees with the results from previous theoretical studies and recent MD simulations for several different types of metallic liquids (*4*). The temperature at which the measured viscosity departs from Arrhenius behavior is labeled in Figure 2.a as $T_{coop}$. While the departure is gradual and difficult to determine directly, it becomes clearer in the residuals from a linear fit (insert to figure). The physical meaning of $T_{coop}$ is intriguing; from MD simulations (*4*) it corresponds to the temperature at which flow first becomes cooperative.

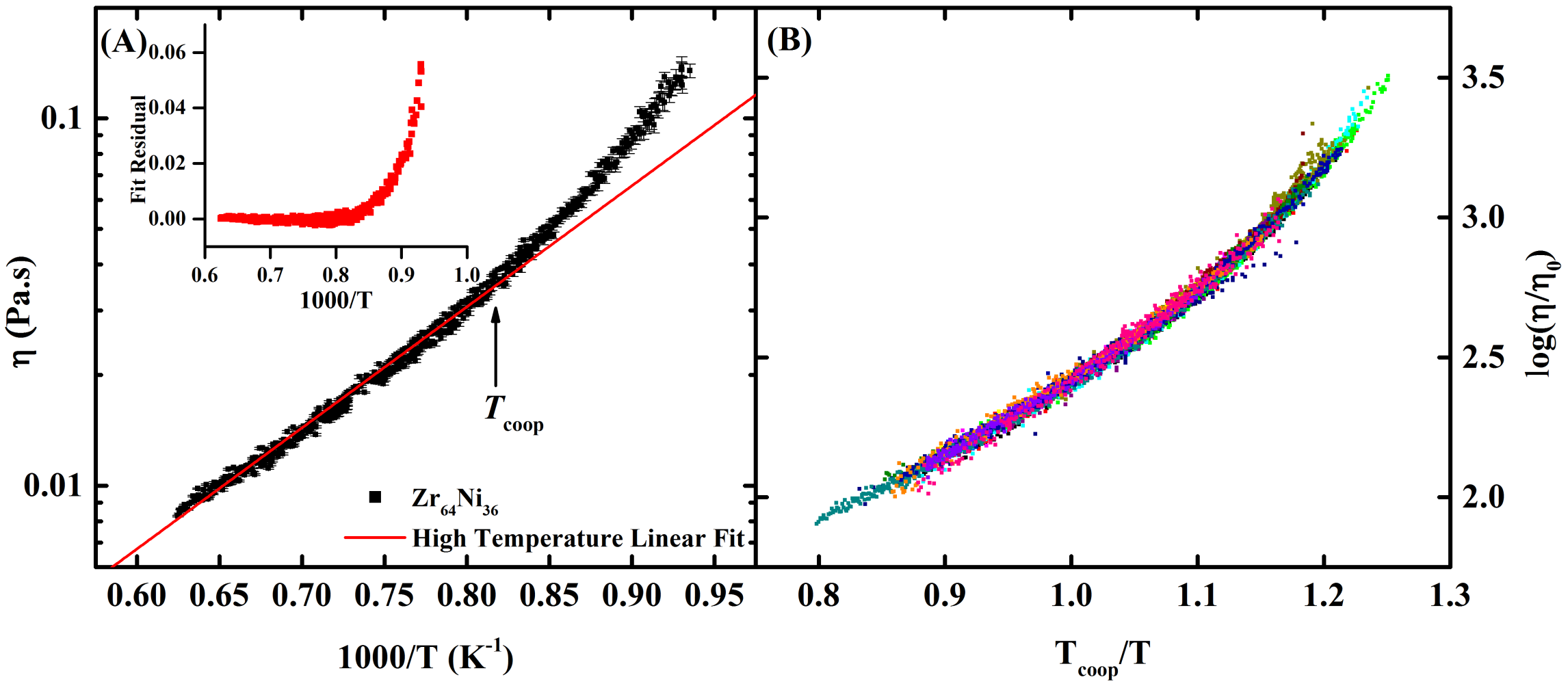

**Figure 2** – (a) Typical example of the behavior of log(η) at high temperature, showing a departure from Arrhenius behavior on cooling below $T_{coop}$. (b) Scaled universal curve for all measured viscosity data.

The MD simulations revealed a universal curve for the ratio of the Maxwell relaxation time for viscosity and the time required to change the local coordination number in a cluster by one unit, by scaling the temperature with the temperature corresponding to the onset of dynamical cooperativity (defined there as $T_A$). Those results suggest that the measured viscosity data could be scaled by $T_{coop}$. As is verified in Figure 2.b. by adopting two material-dependent parameters for scaling, $T_{coop}$ and $\eta_o$, all of the data can be collapsed into a unique curve that describes the temperature dependent viscosity of all of the liquids studied. To construct this curve, the scaling temperature, $T_{coop,}$ was determined for each liquid as illustrated in Figure 2.a, defined as the temperature below which the viscosity became non-Arrhenius. The value of $\eta_o$ for each liquid was then adjusted to collapse the data along the vertical axis. For all liquids studied, on average $\eta_o \approx nh$, where $n$ is the particle density and $h$ is Planck's constant. Exact values and comparisons to $nh$ can be found in Extended Data Table 2.

**Functional Form of the Universal Curve**

While it is shown that the measured data can lie on a universal curve when properly scaled, the temperature range is small. An immediate consequence of the observed data collapse is that, at intermediate temperatures, $\eta = \eta_o F(T/T_{coop})$. Here, $F(z)$ is a universal function common to all of the metallic liquids studied (i.e., one exhibiting no adjustable parameters). To examine whether the scaling holds over a larger temperature range, measured viscosity data near $T_g$ for the strong metallic glass-forming liquid Vit 106a (*8*) were combined with the high temperature data reported here for the same liquid. However, since the scaling in Figure 2.b is empirical, extending it to lower temperatures requires knowledge of the functional form of the universal curve. Many earlier proposed expressions for the viscosity exist, but it has not been previously possible to extensively test them in metallic liquids over the wide temperature range that is possible here, extending from above the melting temperature to the glass transition temperature. The combined Vit 106a data, then, constitute a benchmark for testing these expressions and for identifying the one (if any) that describes the universal behavior. The fits to the Vit 106a data for some of the better-known expressions are shown in Figure 3. These are (i) the commonly used Vogel-Fulcher-Tammann (VFT) equation (*9*), (ii) the recently proposed Mauro-Yue-Elliston-Gupta-Allan (MYEGA) equation (*10*), (iii) a relation derived within the Cohen-Grest free volume model (CG) (*11*), (iv) the avoided critical point theory (KKZNT) (*5-7*), (v) a cooperative shear model (DHTDSJ) (*12*), and (vi) a parabolic kinetically constrained model (EJCG) (*13, 14*). Since the EJCG expression is only valid up to its onset temperature $T_o$, leaving higher temperature behavior undefined, a new variant (BENK) is introduced here, for which the constant Arrhenius type barrier is augmented to give to an expression that emulates the avoided critical point theory when $T$ is replaced by $1/T$. Its crossover temperature $\tilde{T}$ is similar to $T^*$ in the KKZNT expression. These equations are defined in Table 1 and the corresponding optimal parameters are listed in Extended Data Table 1.

**Table 1**

Tested Fitting Functions for Viscosity

| | |
|---|---|
| **Vogel-Fulcher-Tammann (VFT)** | $log\eta = log\eta_0 + \frac{D^* T_0}{T - T_0}$ |
| **Configurational Entropy (MYEGA)** | $log\eta = log\eta_0 + \frac{K}{T}\exp\left(\frac{C}{T}\right)$ |
| **Free Volume (CG)** | $log\eta = log\eta_0 + 2B / \left(T - T_0 + \sqrt{(T - T_0)^2 + CT}\right)$ |
| **Avoided Critical (KKZNT)** | $log\eta = log\eta_0 + \frac{1}{T}\left(E_\infty + T^* B[(T^* - T)/T^*]^z \Theta(T^* - T)\right)$ |
| **Cooperative Shear (DHTDSJ)** | $log\eta = log\eta_0 + \frac{W_0}{kT}\exp\left(-\frac{T}{T_W}\right)$ |
| **Parabolic (EJCG)** | $log\eta = log\eta_0 + J^2\left(\frac{1}{T} - \frac{1}{T_0}\right)^2, T < T_{FitMax} \le T_0$ |
| **Modified Parabolic (BENK)** | $log\eta = log\eta_0 + \frac{E}{kT} + J^2\left(\frac{1}{T} - \frac{1}{\tilde{T}}\right)^2 \Theta\left(\tilde{T} - T\right)$ |

$\Theta(x)$ is the Heaviside function (i.e., $\Theta(x > 0) = 1$ and $\Theta(x < 0) = 0$).

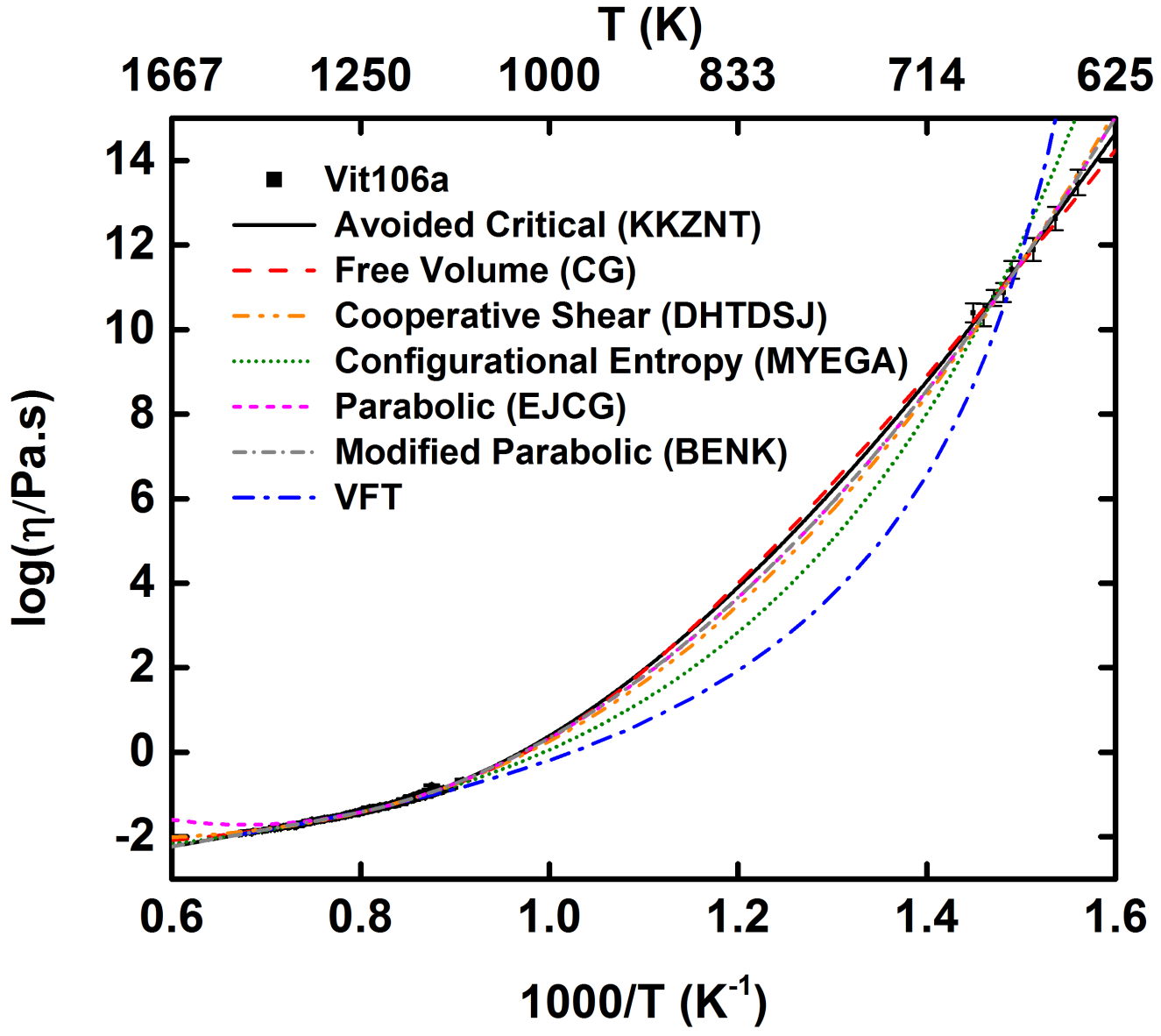

**Figure 3** – Comparison of the fits to the Vit 106a data for the expressions listed in Table 1. The fit parameters can be found in Extended Data Table 1.

All of these expressions fit the high-temperature viscosity data reasonably well, although most do not fit the entire temperature range. In particular, the most-commonly used VFT, and, to a lesser degree the MYEGA, expressions are both in poor agreement with the slope of the data near $T_g$ (672K). Further, since MYEGA does not tend to an Arrhenius form at high temperature, as is predicted by the MD simulations (*4*) and observed in the other models tested, the fit values for $\eta_o$ are much larger than from fits to the other models and from the scaled data shown in Figure 2.b. The CG, DHTDSJ, BENK, and KKZNT models all fit the data over the entire temperature range. The DHTDSJ expression and to a lesser extent the CG form are, however, not consistent with a sharp cross-over to a high temperature Arrhenius type behavior.

We now discuss a particular approximate expression to the universal function $F(z)$ appearing in our data collapse. In earlier studies, the avoided critical point expression (KKZNT) was shown to fit the viscosity data for many non-metallic liquids (*5*-*7*), albeit with five fitting parameters, compared to the three or four parameters for the other models investigated here. Earlier considerations suggested that some of those parameters are fixed. Bolstered by theory, empirical tests (*5*-*7*) yielded an exponent $z \approx 8/3 \pm 1/3$. The KKZNT expression includes an "avoided critical point temperature" $T^*$, which like and $T_A$ in the MD results and the scaling temperature $T_{coop}$ introduced here, corresponds to the onset temperature for dynamical cooperativity. Based on theoretical considerations (*7*), $T^*/T_l = 1.08$ for idealized liquids, where $T_l$ is the liquidus, or melting, temperature - a tendency that is on average, is obeyed, but with significant spread. The universal curve further constrains the KKZNT expression, such that $T_{coop}$ and $\eta_o$ are the only remaining free parameters, with the values of the other parameters (now constants) determined by the fit to Vit 106a. The expression may be written as $\eta = \eta_0 \exp(E/kT)$ with a free energy barrier $E = E_\infty + T_{coop}(bT_r)^z \Theta(T_{coop} - T)$, where $\Theta(x)$ is the Heaviside function and the "reduced temperature" is $T_r \equiv (T_{coop} - T)/T_{coop}$, $E_\infty = 6.466 T_{coop}$, $b = 4.536$, $z = 2.889$. As shown in Figure 4.a, with $T_{coop}$ and $\eta_o$ as the only free parameters this expression gives an excellent fit to all of the high temperature viscosity data shown in Figure 1. The values for $T_{coop}$ and $\eta_o$ obtained from the fits are listed in Table S2 of the supplementary information. Since these fit values are identical to within statistical error to those obtained from the empirical scaling procedure used to find the universal curve for the high temperature data (Figure 2.b) (and since as will be shown soon implicitly $T_{\rm coop} \approx 2.02T_{\rm g}$ and $\eta_{\rm o} \approx nh$), the KKZNT fits are essentially parameter free. As shown in Figure 4.b, this same constrained expression also fits data obtained by other investigators over a wide range of metallic glass families. The reason for the small deviation in the high temperature data for Vit 1 (*12*) is unclear; no similar deviation was observed in our measured data for other liquids. Such good agreement for a range of different metallic liquids provides a striking demonstration of the validity of the KKZNT expression as an approximate functional form for the universal curve. However, the possibility of other reasonable approximate forms is not ruled out. For example, the BENK expression (Table 1) also gives good agreement. The precise functional form for $\eta = \eta_{\rm o}F(T/T_{\rm coop})$ is a matter for future theoretical studies. However, the KKZNT gives sufficiently good agreement with the experimental data over a very wide temperature range that it can be used here to examine the nature of the experimental scaling parameters.

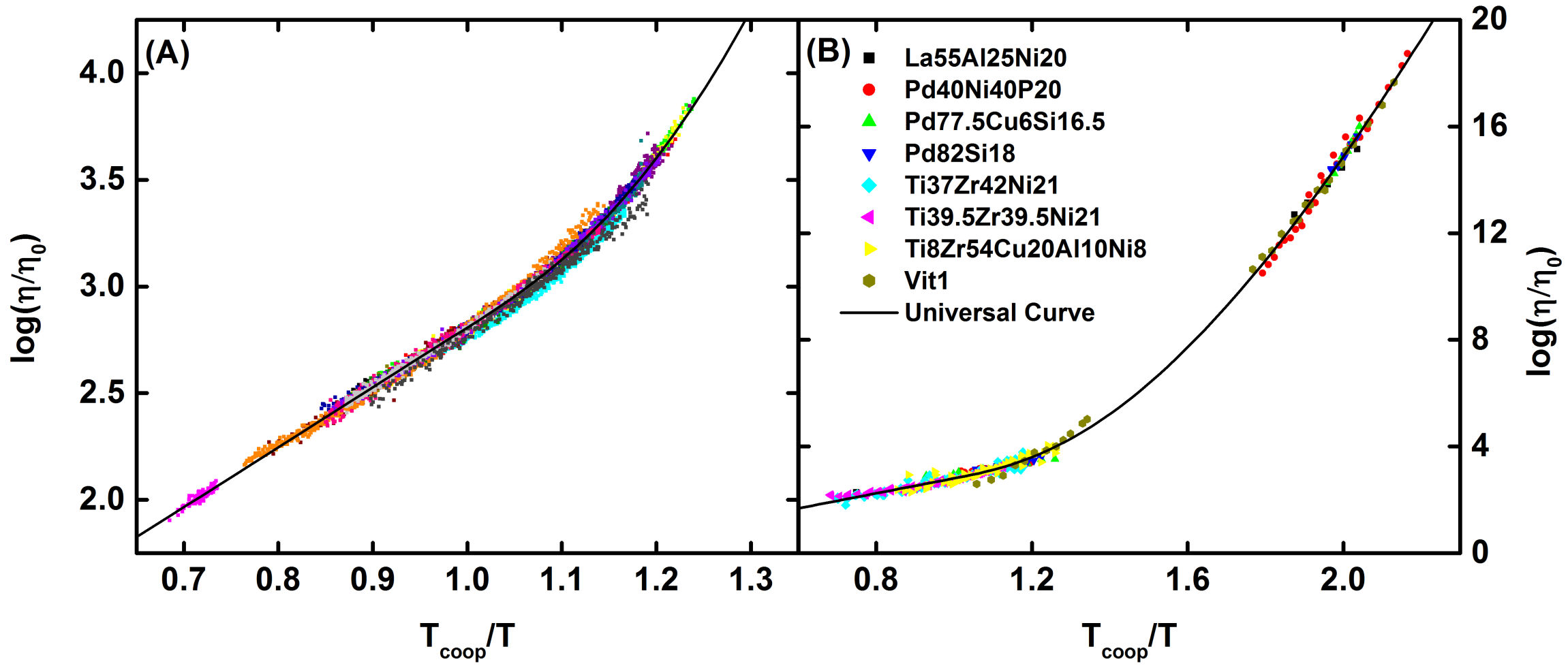

**Figure 4** – (a) Collapse of viscosity data from Figure 1 onto a universal curve assuming the avoided critical point form (KKZNT, black curve). (b) Data collapse of measurements reported in this work and for literature data for additional metallic glass-forming liquids. See Extended Data Table 2 for references to the data for the additional liquids.

Figure 5 shows the correlation between the values of $T_{coop}$ and $T_g$ for the glass-forming liquids. The $T_g$ values were obtained by us and by others from differential scanning calorimetry measurements, using a range of heating rates from 10 to 40 K per minute. The fit line shows that $T_{coop}/T_g = 2.02 \pm 0.015$. This correlation is remarkable, and suggests a deep connection between the onset of cooperative dynamics in the liquid and the dynamical slowing down at the glass transition temperature. It is also the reason that $T_g$ remains a useful scaling temperature.

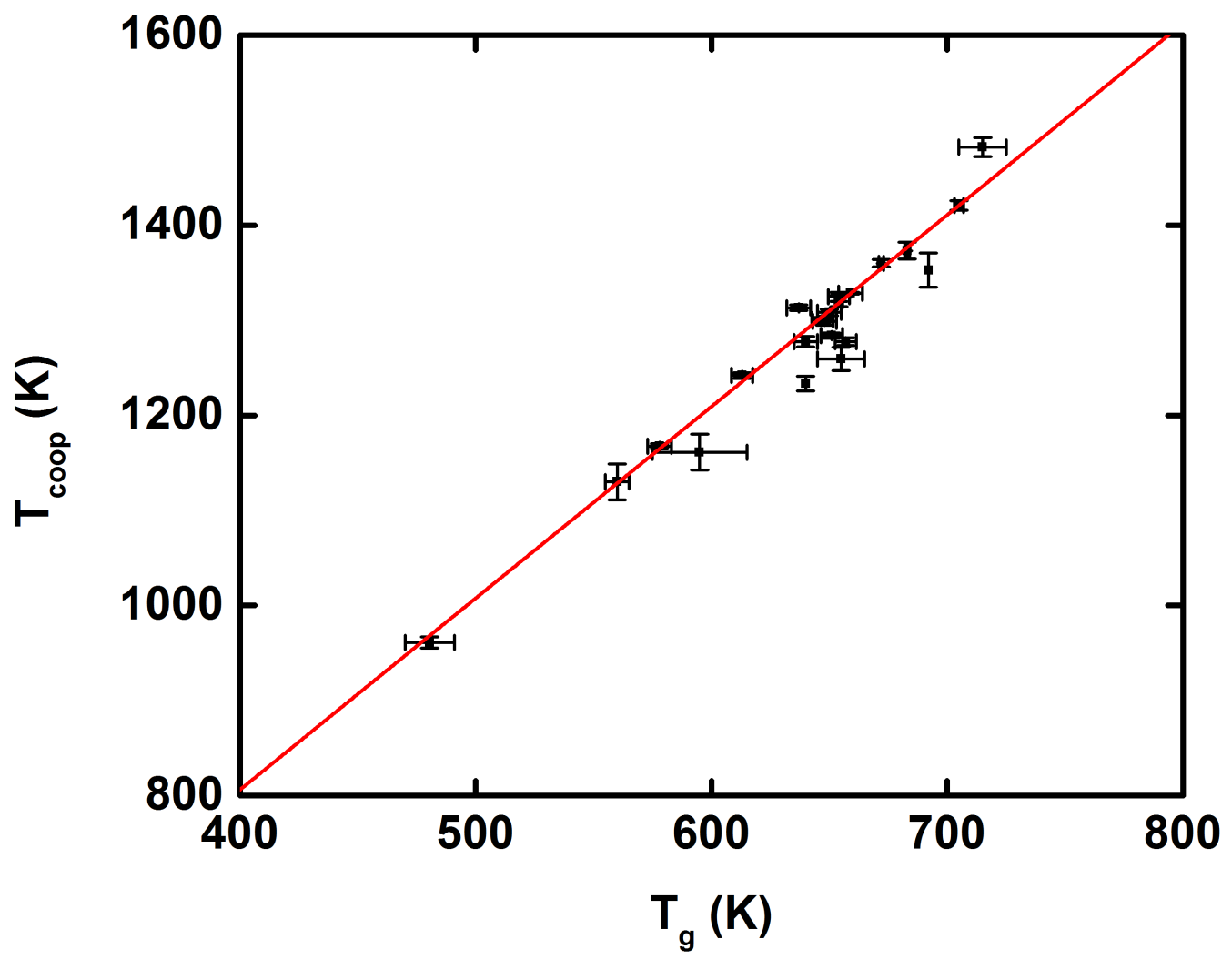

**Figure 5** – Comparison between the experimentally measured glass transition temperature, $T_g$, and $T_{coop}$ from the fits to the high temperature viscosity data. The solid line is a fit to the data, giving $T_{coop} = (2.02 \pm 0.015)T_g$.

This correlation also suggests novel approaches for the search for good glass forming liquids, at least for metallic glasses. For example, assuming the Turnbull criterion of good glass formability when $(T_g/T_l)$ is large (*15*), makes it possible to assess trends in glass formability from liquid data alone, without actually forming a glass and measuring $T_g$. A study of the change in $T_{coop}$ values with the chemical composition of the liquid would show whether such a survey is practically useful.

**Further Discussion of the Scaling Parameters**

As discussed, the scaling parameters $\eta_o$ and $T_{coop}$ can be obtained from fits to the KKZNT theory. But it is important to underscore that they can also be determined empirically by collapsing all of the experimental data onto a universal curve, making them fundamentally theory independent. Several interesting points emerge from an examination of the values obtained for the parameters.

The values for the extrapolated high-temperature viscosity, $\eta_o$, suggest that there may exist a universal high-temperature limit of the viscosity. For the liquid metals studied here, this is *on average* equal to *nh* ( ), where  is Plank's constant and *n* is the particle density per unit volume. Such typical values for $\eta_0$ have been predicted many times, first by Eyring (*16*) from a reaction rate theory argument. It may have an even deeper significance, however, that extends beyond that of liquid metals. Fundamental lower limits on the viscosity are discussed widely in various contexts, with recent interest (*17*) driven by predictions from string theory and holographic dualities, which were compared with measurements at the Relativistic Heavy Ion Collider (*18*).

The strong correlation between $T_{\text{coop}}$ and $T_g$ ($T_{\text{coop}} \sim 2.02\ T_g$) for all of the metallic-glass-forming liquids examined supports a long-held belief from other complementary approaches that glass formation might be a consequence of a high temperature transition crossover. Theories of this crossover include an avoided critical point (*5-7*), a random first order transition (*19, 20*), and mode coupling theories (*21*), among others (e.g. (*22*)). In the avoided critical point theory $T_{\text{coop}}$ corresponds to the transition temperature of the supercooled liquid in an idealized template – a transition that is avoided by frustration. There have been previous experimental hints of non-trivial dynamics associated with a viable dynamical crossover temperature, $T_{\text{cross}}$ (above $T_g$). These include (a) the appearance (at $T < T_{\text{cross}}$ ) of short time, or β, relaxation processes accompanying the primary, or α, relaxation rates that are the focus of this work (*23*), (b) the broadening of relaxation times about these two principal processes (typically this broadening is manifest in response functions that have a stretched exponential behavior) (*24, 25*), (c) nonuniform dynamics in different spatial regions (dynamical heterogeneities) (*26*), (d) violation of the Stokes-Einstein relation (*27*), and (e) decoupling of translational and rotational diffusivity (*28*), and (f) phonon localization (*4*). These phenomena appear and are strongly indicative of transformations that have an onset above $T_g$, yet at temperatures that are lower than 2 $T_g$ in most studied non-metallic liquid systems. It is also important to note that $T_{coop}$ is much above the other predicted dynamic crossover temperature (*22*) and the mode-coupling temperature (*29*).

**Emerging Questions**

The results presented here raise several questions. For example, what is the origin of the observed universal behavior and the connection between $T_g$ and $T_{coop}$? It has long been known that supercooled metallic liquids veer towards locally preferred low energy icosahedral structures (*5-7, 30-32*), a tendency that other liquids generally do not share. This general propensity towards locally preferred structures lies at the origin of the avoided critical point model (*5-7*). MD simulations show that on decreasing the temperature below $T_{coop}$ metallic glass forming liquids progressively develop more pronounced icosahedral order, with a length scale of interpenetrating icosahedral cluster networks that monotonically increases until they percolate throughout the entire system near $T_g$ (see, e.g., Figure 7 of Ref. (*33*)). This is in agreement with predications from avoided critical point theory (*5-7*), and could be the source of the connection between $T_g$ and $T_{coop}$ that is experimentally observed here. Additionally, over the ensemble of metallic liquids that were examined, $T_{coop}/T_l \approx 1.075 \pm 0.188$ (where $T_l$ is the liquidus temperature), consistent with estimates suggested by the avoided critical theory (*7*). However, it should be emphasized that local order need not be icosahedral for theories of an avoided critical point nature to be valid. The local structures of Pd-Si, for example, are likely not strongly icosahedral, while some liquids that have more definitive icosahedral order and still fit the universal curve (e.g., Zr-Pt) deviate more from the predictions of the scaling parameters $\eta_o$ and $T_{coop}$ than do others.

A second question is what these results imply about liquid fragility. If $\eta_o$ were truly independent of temperature for liquid metals, as suggested from the data presented, and if $T_{coop}/T_g$ were truly constant, then the fragility index, defined as $m \equiv (\partial \log_{10}\eta/\partial(T_g/T))_{T=Tg}$, would be the same for all liquids. This is very unlikely, however. Evidence for fragility exists not only in the dynamical properties (where typically an assumed VFT form for $\eta$ is invoked), but also in thermodynamic properties (*34, 35*) and rate of structural ordering (*3, 36*). So, where did fragility go? The clue is in the nature of the deviation of the $T_{coop}/T_g$ ratio found for different liquids as well as remnant small deviations of the viscosity data from our universal collapse. Towards this end, it would be useful to see how the fragility index $m$ varies with $T_{coop}/T_g$ for all of the liquids studied. For metallic liquids fragility is frequently determined from calorimetric measurements(*37*) in addition to or when viscosity data are unobtainable. However, while this gives reliable values of $m$ for good glass-formers, the values for marginal glass forming liquids are scarce and often unreliable. This makes it impossible at present to examine the correlation for all of the liquids studied. Instead, figure S1 compares the ratio $T_{coop}/T_g$ with values of $m$ reported (*8, 38-40*) for the good glass-forming liquids studied here with. The $m$ values range from about 32 to 75; for comparison the $m$ values for $SiO_2$ and o-terphenyl are 20 and 81, respectively. While there are substantial disparities in the values of $m$ that different groups have reported for any single metallic liquid, the average reported fragility index values clearly increase with $T_{coop}/T_g$, in agreement with a structural origin of fragility (see (*36*)). However, as seen from the steepness of the slope in Figure S1, the large variations in the values of $m$ evaluated just above $T_g$ are not similarly reflected in the $T_{coop}/T_g$ ratio. The origin of this enigmatic behavior is not understood.

**Summary and Conclusions**

In summary, a new universality has been observed in the dynamical behavior of liquid metals when the temperature is scaled by $T_{coop}$, which corresponds to the onset of dynamical cooperativity, and when the extrapolated infinite temperature viscosities ($\eta_o$) are properly accounted for. That the glass transition temperature, $T_g$, is strongly correlated with $T_{coop}$ suggests that the cooperative processes that eventually lead to the glass transition are already present in the high temperature liquid. The rapid cooling rates needed for molecular dynamics (MD) studies often raise questions of the validity of comparisons between computed results and experimental data for temperatures near $T_g$. The results presented here, however, show that it is possible to more realistically probe processes associated with the glass transition by comparing computed results with experimental data obtained at high-temperature, where experimental and MD relaxation times are comparable. Taken together, all of these considerations make liquid metals ideal for further fundamental investigations of liquid dynamics.

**Acknowledgements.** We thank Chris Pueblo for some of the DSC measurements. We gratefully acknowledge the authors of Refs. (*24*), (*23*) and F. Mallamace in particular for helpful correspondence. We thank Anup Gangopadhyay, Marios Diemetriou, Jeppe C. Dyra, Eduardo Fradkin, A. L. Greer, Bill Johnson, Flavio Nogueira, and Shmuel Nussinov for useful discussion. The research reported was partially supported by the National Aeronautics and Space Administration (grant NNX10AU19G) and by the National Science Foundation (grants DMR-12-06707, DMR-11-06293). T. Egami was supported by the Department of Energy, Office of Basic Energy Sciences, Materials Science and Engineering Division.

K.F. Kelton supervised the project, coordinated the experimental and theoretical analysis, and prepared the manuscript, in collaboration with the other authors. He, in collaboration with T. Egami, conceived of and demonstrated the possibility of the universal curve. M. Blodgett made the experimental measurements of viscosity. The more detailed analysis of the experimental data was mainly carried out by M. Blodgett with suggestions from Z. Nussinov and K.F. Kelton. Z. Nussinov and T. Egami provided theoretical support. Z. Nussinov examined viable physical content/consequences of the scaling collapse parameters and fits.

## Supplementary Materials:

### Materials & Methods

The samples were prepared from master-alloys that were made by arc-melting on a water cooled hearth in a high-purity argon (99.998%) argon environment. Elements of high purity - 99.9% (Y & Co) to 99.9999% (Cu) - were used to prepare the alloys. When possible, source material was selected for minimum oxygen content (e.g. <10 ppm Zirconium), as this dramatically affects the amount of super-cooling attainable. A Ti-Zr getter was also melted before arc-melting the elements to further reduce the residual oxygen in the atmosphere. The approximately one gram master ingots were melted three times to ensure a homogenous composition; ingots with mass loss greater than 0.05% were discarded. The master ingots were then broken apart and re-melted into samples for the Electrostatic Levitation Studies (ESL); these were in a mass range 40-90 mg.

Samples were then levitated and melted in the high-vacuum containerless environment of the Washington University Beamline ElectroStatic Levitation Facility (WU-BESL). The absence of a container and the high-vacuum environment (~$10^{-7}$ Torr) minimized heterogeneous nucleation, allowing data to be collected from both equilibrium and supercooled liquids. More details of the WU-BESL can be found elsewhere (*41*).

The viscosity was measured as a function of temperature using an oscillating drop method (*42*). The voltage on the vertical electrode was modulated at a frequency that near the $l$ = 2 spherical harmonic mode resonant frequency (typically 120–140 Hz) of the liquid to induce surface vibrations. A high-speed camera (1560 frames per second) was used to capture the shadow of the oscillating sample. After the oscillation was stable, the perturbative voltage was removed and the time-dependent amplitude of the decaying surface harmonic oscillations was measured. The viscosity was determined from the decay time for the oscillation, τ,

$$\eta = \rho R_0/(l-1)(2l+1)\tau$$

where $\rho$ is the density and $R_0$ is the unperturbed radius of the sample. The small magnitude of the viscosity over the measurement range and the low strain rates ensure that shear thinning does not influence the measurements.

**Table S1**

Values of Parameter from Fits to Vit106A, as Shown in Figure 2 and Table 1 of the Main Text.

| Fitting Equation | $log_{10}\eta_0$ | Other Parameter Values |
|---|---|---|
| **Vogel-Fulcher-Tammann (VFT)** | -3.48 | $D^*$=5.60, $T_0$=575 |
| **Configurational Entropy (MYEGA)** | -2.80 | K=237, C=2.49x$10^3$ |
| **Free Volume (CG)** | -2.67 | B=873, C=58.6, $T_0$=954 |
| **Avoided Critical (KKZNT)** | -4.50 | $E_\infty$=3819, $T^*$=1360,B=34.3,z=2.889 |
| **Cooperative Shear (DHTDSJ)** | -2.11 | $W_0$=1.36 x$10^5$, $T_W$=247 |
| **Parabolic (EJCG)** | -1.72 | $J^2$=1.97 x$10^7$, $T_0$=1476 |
| **Modified Parabolic (BENK)** | -4.64 | E=4.01x$10^3$, $J^2$=1.96x$10^7$, T~=1285 |

**Table S2**

Scaling Parameters $\eta_0$ and $T_{coop}$ and their Relation to the Predicted High Temperature Viscosity Limit ( and the Glass Transition Temperature,.

| Composition | *Log(nh)* | *Log*$\eta_0$ | $T_{coop}$ | $T_g$ |
|---|---|---|---|---|
| $Cu_{50}Zr_{45}Al_5$ | -4.441 | -4.603 | 1308.37 | 650 |
| $Cu_{50}Zr_{50}$ | -4.448 | -4.598 | 1283.72 | 651 (*43*)* |
| $Cu_{60}Zr_{20}Ti_{20}$ | -4.386 | -4.731 | 1300.54 | 647 |
| $Ni_{75}Si_{25}$ | -4.294 | -4.148 | 1071.65 | - |
| $Ti_{40}Zr_{10}Cu_{30}Pd_{20}$ | -4.409 | -4.613 | 1298.68 | 648 |
| $Ti_{40}Zr_{10}Cu_{36}Pd_{14}$ | -4.402 | -4.635 | 1277.56 | 640 |
| Vit106 [$Zr_{57}Cu_{15.4}Ni_{12.6}Al_{10}Nb_5$](*8*) * | -4.480 | -4.445 | 1373.14 | 683 (*8*) * |
| Vit106A [$Zr_{58.5}Cu_{15.6}Ni_{12.8}Al_{10.3}Nb_{2.8}$](*8*) * | -4.480 | -4.502 | 1360.00 | 672 (*8*) * |
| $Y_{68.9}Co_{31.1}$ | -4.573 | -4.398 | 1130.05 | 560 |
| $Zr_{59}Ti_3Cu_{20}Ni_8Al_{10}$ | -4.497 | -4.516 | 1320.34 | 652 |
| $Zr_{60}Ni_{25}Al_{15}$ | -4.502 | -4.501 | 1420.80 | 698 |
| $Zr_{62}Cu_{20}Ni_8Al_{10}$ | -4.501 | -4.452 | 1324.65 | 654 |
| $Zr_{64}Ni_{36}$ | -4.485 | -4.247 | 1223.38 | - |
| $Zr_{70}Pd_{30}$ | -4.525 | -4.394 | 1328.56 | 659 |
| $Zr_{75}Pt_{25}$ | -4.500 | -4.404 | 1550.02 | - |
| $Zr_{76}Ni_{24}$ | -4.522 | -4.246 | 1161.26 | 595 |
| $Zr_{80}Pt_{20}$ | -4.514 | -4.361 | 1481.96 | 715 (*44*) * |
| **Literature Data** | | | | |
| $La_{55}Al_{25}Ni_{20}$ (*12, 45*) * | -4.642 | -5.018 | 966.379 | 481 (*46*) * |
| $Pd_{40}Ni_{40}P_{20}$ (*12, 45, 47*) * | -4.320 | -4.820 | 1170.61 | 578 (*45*) * |
| $Pd_{82}Si_{18}$ (*48, 49*) * | -4.411 | -4.871 | 1277.43 | 657 (*50*) * |
| $Pd_{77.5}Cu_6Si_{16.5}$ (*12*) * | -4.406 | -4.867 | 1313.05 | 637 |
| $Ti_{37}Zr_{42}Ni_{21}$ (*51*) * | -4.486 | -4.325 | 1176.53 | - |
| $Ti_{39.5}Zr_{39.5}Ni_{21}$ (*52*) * | -4.482 | -4.181 | 1143.67 | - |
| $Ti_8Zr_{54}Cu_{20}Al_{10}Ni_8$ (*53*) * | -4.491 | -4.339 | 1259.29 | 655 |
| Vit1 [$Zr_{41.2}Ti_{13.8}Cu_{12.5}Ni_{10}Be_{22.5}$] (*12*) * | -4.468 | -3.2986 | 1241.87 | 613 (*12*) * |

*References to viscosity and calorimetry data obtained by other investigators.

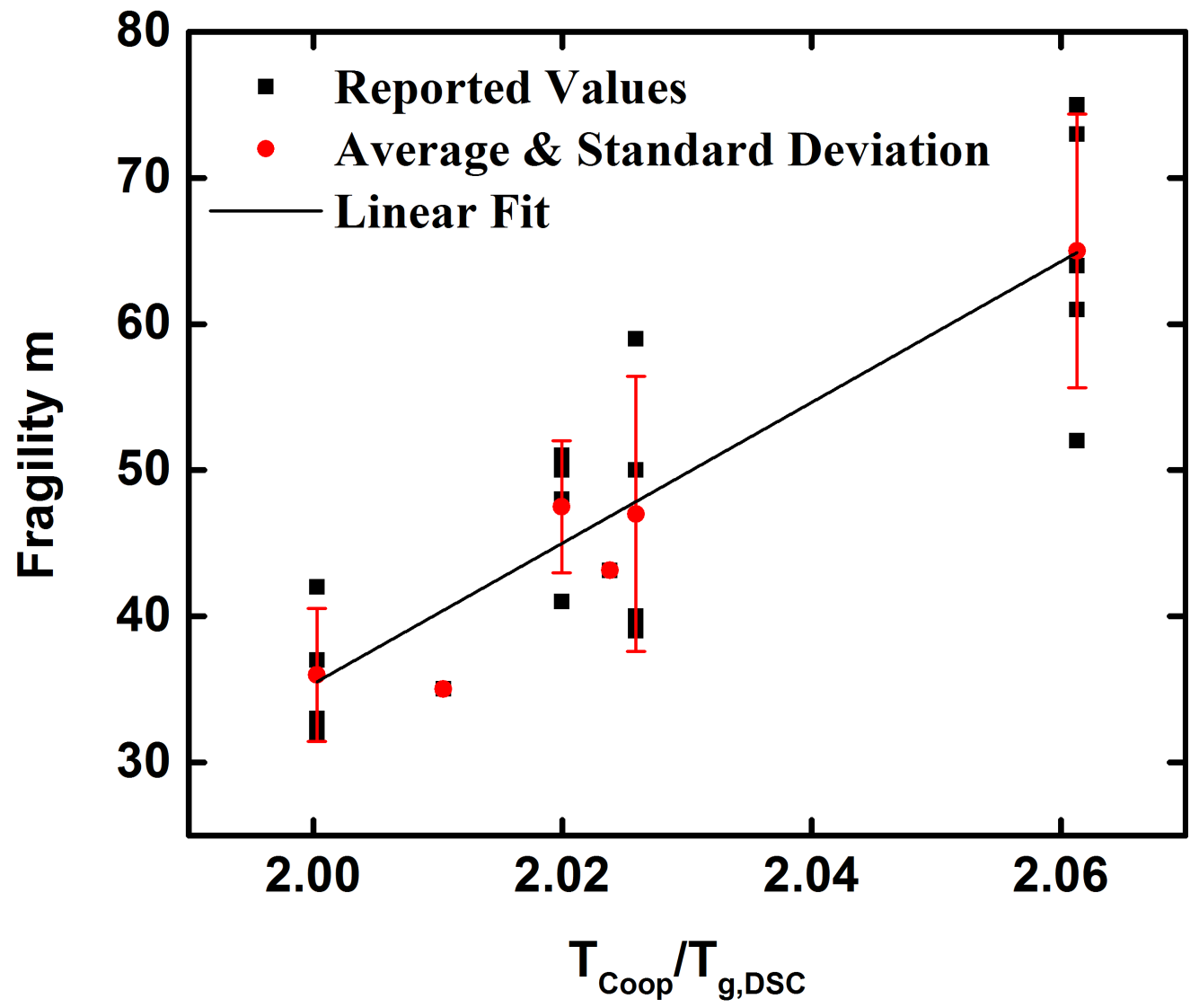

**Figure S1** – Published fragility (*m*) data ( ) versus $T_{coop}/T_g$. The filled red circles ( ) represent the average values; the error bars reflect the standard deviation. The large scatter in the reported *m* values for these bulk metallic glasses, as well as the lack of data for marginal glass-formers, reflects the difficulty in measuring *m*.